\documentstyle[12pt,prb,aps]{revtex}
\begin{document}
\draft
\title{ Quantum weak chaos in a degenerate system}
\author{V. Ya. Demikhovskii\footnote{demi@phys.unn.runnet.ru}, 
D. I. Kamenev\footnote{kamenev@phys.unn.runnet.ru},}
\address{Nizhny Novgorod State University, Nizhny Novgorod,
603600, Russia}
\author{G. A. Luna-Acosta\footnote{gluna@sirio.ifuap.buap.mx}}
\address{Instituto de Fisica,
Universidad Autonoma de Puebla, Puebla, Pue. 72570,
Apdo. Postal J - 48
Mexico}
\maketitle
\bigskip
\begin{abstract}
Quantum weak chaos is studied in a perturbed
degenerate system --- a charged particle interacting with a
monochromatic wave in a transverse magnetic field. The evolution
operator for an arbitrary number of periods of the external field is
built and its structure is explored in terms of the QE (quasienergy
eigenstates) under resonance condition (wave frequency $=$ cyclotron
frequency) in the regime of weak classical chaos. The new
phenomenon of
diffusion via the quantum separatrices and the influence of chaos on
diffusion are investigated and, in the quasi classical limit,
compared with
its classical dynamics. We determine the crossover from purely quantum
diffusion to a diffusion which is the quantum manifestation of classical
diffusion along the stochastic web.
This crossover results from the non-monotonic dependence of the
characteristic localization length of the QE states on the wave amplitude.
The width of the quantum separatrices was computed and compared with the
width of
the classical stochastic web. We give the physical parameters which can be
realized experimentally to show the manifestation of quantum chaos in
nonlinear acoustic resonance.
\end{abstract}
\bigskip


\section{Introduction}
The problem of quantum chaos in intrinsically
degenerate systems possesses a number of interesting properties. The
KAM theorem in these systems is not applicable \cite{comment1} and in
certain models an arbitrary small
perturbation is sufficient to induce an infinite stochastic
web in phase space. The character of the web is determined by the type of  perturbation. When the system is perturbed by a monochromatic wave, the
web width exponentially decreases with increasing action $I$ and the
motion is practically localized.\cite{Zaslavsky} If
the perturbation has the form of periodic $\delta$ - impulses, the web
width is constant and the particle, traveling along the web, can diffuse
to infinity \cite{Zaslavsky} (see also Refs. \cite{D1,Z2}). In both
cases the web in ($p$,$x$) phase space has a crystalline or
quasicrystalline structure. The chaotic motion in these systems
has been termed weak chaos since chaos occupies only a small portion
of phase space ( See e.g., Ref. 2).
In contrast, in strong ( or global ) chaos, the web structure disappears
and most of the phase space is filled with chaotic orbits.
The peculiar resonance structure and the appearance of an 
infinite stochastic web make degenerate systems very attractive
objects in which to study quantum manifestations of chaos.

The problem of weak and strong chaos has been mainly
explored within the context of the kicked harmonic oscillator.
\cite{kicked_oscillator,Dana} It has been shown that the time of classical
description of quantum averages is considerably longer for weak than for 
strong chaos. The role of the symmetry of the
quasienergy (QE) functions was also analyzed in Ref. 
\cite{kicked_oscillator}. 
It was found that under certain
conditions, quantum diffusion within the stochastic web was
truncated by quantum
interference effects \cite{Dana}, similar to the case of strong chaos
in the kicked rotor \cite{rotor}. The problem of quantum
chaos on the stochastic web has also been studied intensively in
recent years within the framework of the generalized kicked Harper
model. \cite{KH1,KH2} Another simple and yet important (especially to
solid state physics)  example of a degenerate
system is that of a charged particle moving in a uniform magnetic
field and interacting with a monochromatic wave, propagating
perpendicularly to the magnetic field direction under the condition of
cyclotron resonance. This 2D problem  is
tantamount to the 1D harmonic oscillator in a wave field.\cite{D1,D2} In
our previous work \cite{1} we studied this system quantum mechanically, 
focusing on the
resonance approximation, which appears as the first order perturbation 
for the Floquet Hamiltonian. In the quasiclassical limit the dynamics 
in the resonance approximation is globally regular in phase space. The 
structure of the Floquet spectrum and
the QE (quasi-energy) eigenfunctions for the exact and near resonance cases 
were obtained and related to the classical phase space structure. The
evolution of various representative initial states was investigated
and the close connection between classical and quantum dynamics at the
cyclotron resonance was demonstrated. In Ref. \cite{2} it was shown
that the boundaries of the quantum cells act as dynamical barriers to
the probability flow. In the quasiclassical limit the dynamical
barriers were found to correspond to the separatrices in classical
phase space and tunneling  through the  ''quantum
separatrices '' was explored numerically.

In the present work we study the dynamical effects of chaos
on the above mentioned system under the
condition of  cyclotron resonance $\omega=\omega_c$
($\omega$ and $\omega_c$ are the wave and cyclotron frequencies, 
respectively). This
model seems to be more closely related to experimental realizations in
solid state physics than the kicked system. The evolution operator for
an arbitrary number of periods of the external field is built and its
structure is explored in terms of the QE eigenstates under the conditions
of weak chaos.
The structure of the evolution
operator matrix  is more complex than the more typical
band-like matrix structure.
Thus, the usual diagnostics of quantum chaos predicted by band random
matrix theory \cite{RMT1,RMT2} do not work in the case of weak
chaos because only a small number of QE eigenstates is affected by
the perturbation. A new phenomenon of diffusion over the quantum
separatrices and the effect of weak chaos on the diffusion
are investigated here and compared in the quasiclassical
limit with the dynamics in classical phase space. The crossover
from purely quantum diffusion to a diffusion which in the
quasiclassical limit corresponds to classical diffusion
within the stochastic
web, is determined. The width of the
quantum separatrices is computed and compared with the width of the
classical stochastic web.

The paper is organized as follows. In Sec. II the basic model is
introduced. Also in Sec. II, the structures of the nonstationary
Schr\"odinger equation and the evolution operator in $\hat H_0$
representation are discussed. In Sec. III the properties of the QE
eigenstates are described. The phenomenon of diffusion via       the
quantum separatrices and the influence of chaos on the diffusion
are investigated numerically in Sec. IV. In Sec.V we draw our
conclusions.

\section{The evolution operator}
The Hamiltonian of a charged particle in a magnetic field interacting
with a monochromatic wave reads \begin{equation}
\label{general_Hamiltonian} \hat H=\frac{(\hat{\bf p}+{e\over c}{\bf
A})^2}{2m}+v_0\cos(kx-\omega t)= \hat H_0+\hat V(x,t), \end{equation}
where $m$ and $e$ are, respectively, the mass and charge of the
particle, $\hat{\bf p}$ is the momentum, $k$ is the wave vector,
$\omega$ is the wave frequency, and $v_0$ is the amplitude of the
perturbation. We choose the gauge of {\bf A}  in the form {\bf A} $=$
(0,H$x$,0) so as to have the magnetic field {\bf H} along the
z-direction and to have the momentum $p_y$ as an integral of motion.
The Hamiltonian is equivalent to a 1D simple harmonic oscillator 
perturbed by a monochromatic wave field.
Hence, the problem is to determine the dependence of the wave function on
only two variables, $x$ and $t$.

It is convenient to expand the state vector in the harmonic oscillator
basis
 \begin{equation}
 \label{Landau_series}
 \psi(x,t)=\sum_n C_n(t)\psi_n(x)\exp(-iE_nt/\hbar),
 \end{equation}
 where $\psi_n(x)$ is the nth eigenfunction of the
 simple harmonic oscillator Hamiltonian
 $\hat H_0$ and $E_n=\hbar\omega_c(n+1/2)$ is the energy of the
 nth Landau level.
 Using (\ref{Landau_series}), the nonstationary Schr\"odinger equation
 \begin{equation} \label{SH_Eq} i\hbar\frac{\partial\psi
 (x,t)}{\partial t}=\hat H\psi (x,t) 
 \end{equation}
 yields a set of differential-difference
 equations for the coefficients $C_n(t)$,
 \begin{equation}
 \label{motion_equations}
 i\hbar\dot C_n=
 v_0\sum_m \left[ V_{n,n+m}^{(1)}\sin (\omega t)+
 V_{n,n+m}^{(2)}\cos (\omega t)\right] C_{n+m}
 e^{-im\omega_c t}.
 \end{equation}
 The matrix elements $V_{n,n+m}^{(1)}$ ($V_{n,n+m}^{(2)}$) describe
 the transitions between the levels of opposite (equal) parity and can
 be expressed via the Laguerre polinomials as \cite{1}
 \begin{mathletters} \begin{equation} \label{odd_matrix_elements}
 V_{n,n+2m+1}^{(1)}=\frac{(-1)^m h^m e^{-\frac h4}}{2^{m+1}
 \sqrt{(n+1)\dots (n+2m+1)}}L_n^{2m+1}(\frac h2) , \end{equation}
 \begin{equation} \label{even_matrix_elements}
 V_{n,n+2m}^{(2)}=\frac{(-1)^m h^m e^{-\frac h4}}{2^{m+1}
 \sqrt{(n+1)\dots (n+2m)}}L_n^{2m}(\frac h2) , \end{equation}
 \end{mathletters} where $h =(ka)^2$ plays the role
of an effective (dimensionless) Planck constant and
 $a=\sqrt{\hbar c/e{\text H}}$ is the magnetic length.
 For $n\gg 1\gg h$
the matrix elements
 can be approximated in terms of the Bessel functions $J_m$ of order $m$
 by \cite{Gradstein} \begin{mathletters}
\begin{equation}
\label{Bessel}
V_{n,n+2m+1}^{(1)}=\frac{1}2\frac{(-1)^m n^{m+1/2}
e^{-\frac{h}4}}{\sqrt{(n+1)\dots (n+2m+1)}}J_{2m+1}(\sqrt{2nh}) ,
\end{equation} \begin{equation} V_{n,n+2m}^{(2)}=\frac{1}2\frac{(-1)^m
n^m e^{-\frac{h}4}} {\sqrt{(n+1)\dots (n+2m)}}J_{2m}(\sqrt{2nh}) .
\end{equation} \end{mathletters}

Since the perturbation is periodic in time, Floquet theory can
be used to describe the time evolution of the system in terms
of the QE spectra $\varepsilon_q$ and the QE eigenfunctions
$\psi_q(x,t)$.
The QE states are the eigenstates of the evolution operator $\hat U$
for one period of oscillation of the external field $T=2\pi/\omega$,
$$ \hat U(T)\psi _q(x,t)=\exp(-\frac{i\varepsilon_qT}\hbar)\psi
_q(x,t), $$ which can be defined by (see,for example, Ref.
\cite{Reichl_book}, p.385)
 \begin{equation}
 \label{QE_decomposition}
 \psi _q(x,t)=\exp(-\frac{i\varepsilon_qt}\hbar)\sum_n C_n^q(t)
 \psi_n(x)=\exp(-\frac{i\varepsilon_qt}\hbar)u_q(x,t),
 \end{equation}
where the functions $C_n^q(t)$ and $u_q(x,t)$ are periodic in time,
$u_q(x,t+T)=u_q(x,t)$.

The coefficients $C_n^q(t)$ are the eigenvectors of the operator $\hat U$ 
in the representation of the Hamiltonian $\hat H_0$ and can be
found  by diagonalizing the corresponding matrix $U_{n,m}$.
The following procedure is one way to obtain the matrix elements
\cite{Reichl_proedure}. Let the evolution operator $\hat U$
act on the initial state
 $C_n^{(n_0)}(0)=\delta_{n,n_0}$;
\begin{equation}
\label{Unm} U_{m,n}(T)C_n^{(n_0)}(0)=U_{m,n_0}(T)=C_m^{(n_0)}(T).
\end{equation} The coefficients $C_m^{(n_0)}(T)$ can be computed
numerically by integration of Eq. (\ref{motion_equations}). They form
a column in the matrix $U_{m,n_0}(T)$. Repetition of this process for
initial states, orthogonal to the previous one,
$C_n^{n'}(0)=\delta_{n,n'},\, n'\ne n_0$, fills the matrix
$U_{m,n_0}(T)$. Diagonalization of the matrix $U_{m,n}(T)$ yields
the eigenvalues $\varepsilon_q$ and the eigenvectors $C_n^q$.
The number of the Landau levels $N$ in Eqs. (2), (4) included in our
computations is equal to the size of the evolution operator
matrix $U_{m,n}(T)$ and, hence, to the number of its eigenstates ---
QE eigenstates.

Once the eigenvalues $\varepsilon_q$ and the eigenvectors
$C_n^q$ are obtained, then we may write
the evolution operator for one period $U_{m,n}(T)$ in the form
\cite{Reichl_book} 
\begin{equation} \label{1_T}
U_{n,n'}(T)=\sum_q C_n^qC_{n'}^{q*}\exp({-i\varepsilon_qT/\hbar}).
\end{equation}
By raising $U_{n,n'}(T)$ to degree $M$ and using the
orthogonality of the eigenvectors $C_n^q$, one can obtain the evolution
operator which propagates the system toward $M$ periods $U_{n,n'}(MT)$ by 
\begin{equation} \label{Green_f} U_{n,n'}(MT)=\sum_q
C_n^qC_{n'}^{q*}\exp(-i\frac{\varepsilon_qMT}\hbar). \end{equation}
Given
$U_{n,n'}(MT)$, the evolution of any initial state $C_n(0)$ can be
computed by using 
\begin{equation} \label{Green_evolution}
C_n(MT)=\sum_{n'}U_{n,n'}(MT)C_{n'}(0). \end{equation} 
The expressions
($\ref{Green_f}$) and (\ref{Green_evolution}) are much more practical
for calculations than the integration of the set of differential
equations (\ref{motion_equations}) , especially in the limit
$t\rightarrow\infty$, because they allow us to obtain the state of the
system at any time $t$ by a simple summation.
There are only two dimensionless parameters determining 
the dynamics of the system, namely the dimensionless amplitude 
of the perturbations $V_0=v_0/\hbar\omega$ and the effective Planck
constant $h$ in the arguments of the matrix elements 
(\ref{odd_matrix_elements}) and (\ref{even_matrix_elements}). 
This is easy to see if we write Eq.(4) in dimensionless form by 
introducing the dimensionless time $\tau=t\omega V_0$. In this form,
the phases of the oscillating terms are given by 
$-m\tau/ V_0$. Thus, the larger the
amplitude of the wave $V_0$ is, the smaller the frequency of the
oscillations becomes and the larger the number of the effective terms
which participate in the dynamics. Consequently, the parameter $V_0$ 
determines
the number of effective terms on the right-hand side of  
Eq. (\ref{motion_equations}) which can be roughly
estimated as $m\ge V_0$. The same estimation yields the bandwidth of 
the evolution operator matrix (\ref{Unm}) or (\ref{1_T}). If
$V_0\ll 1$ the resonant terms with $m=\pm 1$ dominate the dynamics;
they become independent of time since they are being multiplied by
$\sin(\omega t)$ or $\cos(\omega t)$. The other terms oscillate
fast  and can be averaged out. Consideration of only these
time-independent coefficients constitutes what we call the resonance
approximation. As was shown in Refs. \cite{1,2}, the quantum resonance
approximation in the quasiclassical limit in general corresponds to
the classical resonance approximation (see e.g. Ref.
\cite{Lichtenberg}). It is necessary to point out that when $V_0\ge 1$,
the matrix elements $V_{n,n+m}\sim J_m(\sqrt{2nh})$ decrease quickly with
$m$ in the region $m>\sqrt{2nh}$ and this inequality can be treated as
an additional restriction on the bandwidth of the evolution
operator matrix.

As $V_0$ increases, the number of effective terms
increases too. This number was numerically determined
by requiring relatively small changes in the QE spectrum
and in the QE eigenfunctions after including an additional
term to sum (\ref{motion_equations}).
The fluctuations did not exceed the accuracy of the Runge-Kutta method
used for integration of Eq. (\ref{motion_equations}). The numerically
determined number of effective terms was found to be of the order of $V_0$
in agreement with the speculations presented above.
The
Runge-Kutta procedure was controlled by the normalization condition
$\sum_n|C_n|^2=1$; the fluctuations of this value were smaller than
$10^{-4}$.

\section{The separatrix quasienergy eigenstates}
In this section we discuss the structure of the QE states which
determine the dynamics via Eqs. (\ref{Green_f}) and
(\ref{Green_evolution}). We first consider the resonance approximation
which will be the starting point for the investigation of quantum
chaotic effects in next order approximations. \cite{1,2} The equation
for the QE eigenstates in the resonance approximation can be obtained
by putting the QE eigenfunction in
the form (7) into Eq. (\ref{motion_equations})
 and keeping only time-independent
(resonance) terms. Thus, the set of differential-difference
equations (\ref{motion_equations}) is transformed into the
set of algebraic equations
\begin{equation}
\label{algebraic}
E_qC_n^q=V_0(V_{n,n+1}C_{n+1}^q+V_{n,n-1}C_{n-1}^q),
\end{equation}
where $E_q=\varepsilon_q/\hbar\omega$ is the dimensionless
quasienergy. This is an eigenvalue problem for the Floquet Hamiltonian.
\cite{2} Eq. (12) is similar to the Harper equation in the symmetric
gauge of the vector potential ${\bf A}$ with periodic off-diagonal
modulation of the matrix elements.\cite{Harper} In our case the off-diagonal
modulation is nonperiodic. The dependence of the matrix
elements $V_{n,n+1}$ on the Landau number $n$ is shown in the upper
part of Fig. 1. Due to oscillations of $V_{n,n+1}$ and $V_{n,n-1}$
with $n$, the Floquet Hamiltonian matrix for determining the QE
eigenstates (\ref{algebraic}) has a cell structure; the boundaries of
the cell are given by the zeroes of the Bessel function (c.f. Eq.
(\ref{Bessel})). One can easily show that at small values of $E_q$
( $V_0\ll 1$) the Floquet Hamiltonian matrix (\ref{algebraic})
can be obtained from the Floquet matrix (\ref{1_T}). Moreover, as will
be shown below, the cell structure is maintained even for very large
values of $V_0$ ($V_0\sim 10$); an extremely strong perturbation
amplitude is required to destroy the cells entirely. The regions where
the matrix elements $V_{n,n+1}$ are small (one such region is marked
with the rectangle in the plot $V_{n,n+1}(n)$ in Fig. 1 ) can be
referred to as "quantum separatrices" because in the quasiclassical
limit their positions in action $I$ correspond to the positions
of the classical separatrices in  phase space. \cite{1,2} These are given 
by the zeroes of the Bessel function of order 1.

The structure of the QE eigenfunctions
can be understood by characterizing each
one by its center $\bar n_q = \sum_n n|C_n^q|^2$
and its dispersion
$\sigma_q = \left(\sum_n (n-\bar n_q )^2|C_n^q|^2\right)^{1/2}$. The plot of
$\bar n_q$ versus $\sigma_q$ in the resonance approximation is shown
in the left part of Fig. 2 (a); the figures on the right are 
the Poincar\'e surfaces of sections for the classical system with the same
parameters. Each point in the plot $\bar
n_q(\sigma_q)$ corresponds to  two QE eigenstates ($E_q$ and $-E_q$),
due to the symmetry of Eq. (\ref{algebraic}) under the
transformation 
\begin{equation}
\label{simmetry} 
E_q \rightarrow -E_q, \qquad C_n^q \rightarrow (-1)^nC_n^q, 
\end{equation} 
corresponding to
the transformation $x \rightarrow -x$ in Eq. (\ref{QE_decomposition}).
It is seen that almost all the QE eigenfunctions are divided into
groups with practically the same $\bar n_q$ and different $\sigma_q$.
Each group of states belongs to only one resonant cell because the $\bar n_q$
for each QE eigenfunction is situated in the center of the cell and
$\bar n_q\pm \sigma_q$ does not exceed the size of the corresponding
cell (the boundaries of the cells in Fig. 2 (a) are marked with
arrows). It was shown in Ref. \cite{2} that the Husimi function
\cite{Husimi} of a QE state with quasienergy $E_q>0$
is localized in the upper part ($x>0$) of the phase space and the
Husimi function of  a state with $-E_q$ is situated in the lower part
($x<0$). Thus, each row on the left part of Fig. 2 (a) corresponds
to the two symmetrical classical resonance cells shown in the right
part of Fig. 2 (a): the first row is associated with two classical cells
near the point ($x=0,\,p=0$); the second row  with the next
symmetrical classical cells; and so on.
The number of the QE eigenfunctions in an individual cell equals
approximately the number of the Landau states in this cell.

Besides the localized eigenfunctions (arranged in  rows in
Fig. 2 (a)), there are
a small number (3 - 4 percent) of delocalized states which cannot
be assigned to any particular cell; they are represented by the
scattered  points. These QE eigenfunctions have
large dispersions, with $\sigma_q$ exceeding the size of one  cell. 
Hereafter we shall characterize the "localization length" of the 
QE functions by $\sigma_q$.
It was found numerically that the most localized eigenfunctions
correspond to the largest QE  
eigenvalues (in absolute value), while the most delocalized states 
correspond to one of the smallest eigenvalues. Calculations with
other values of the effective Planck constant $h$ have shown that the
number of delocalized QE eigenstates increases with increasing $h$
which indicates that, as will be shown below, these states are of a pure
quantum nature with no classical analogs. A representative
delocalized eigenfunction, the widest one, marked in Fig. 2 (a) with
an arrow, is shown in the lower part of Fig. 1, and its Husimi function
is plotted in Fig. 3.
Note that the eigenfunction as
well as the Husimi function have their maxima in the regions of the
classical separatrices (c.f. Fig. 2 (a)). Thus, the delocalized
states can be identified
as "separatrix eigenstates". The Husimi function in Fig. 3 is
symmetrical with respect to $p\rightarrow -p$ but not symmetrical
under the transformation $x\rightarrow -x$, which turns the Husimi
function corresponding to the eigenvalue $E_q$ into the one
corresponding to $-E_q$ (see Eq. (\ref{simmetry})). The high peaks
in Fig. 3 near the separatrix line $x=0$ result from the slowing of
the motion of the classical particle which, in turn, increases the
probability of finding the particle in this region. The existence of the
fully delocalized eigenstates of the matrix $U_{n,m}(T)$ is a very
interesting, nontrivial feature. Initially ( time $M=0$), the QE
functions, due to their completeness, yield a Kroenecker delta
$\delta_{n,n'}$ in Eq. (\ref{Green_f}).
If $n$ and $n'$ belong to
different cells, then only a small number of the delocalized QE functions provides
the cancellation of  terms in Eq. (10). The condition of completenes (10) (at M=0) 
serves as a good check for our numerical calculations.
At short times
($M\sim 1$) in the resonance approximation, where $V_0\ll 1$,
one may take into consideration only the elements along the first
off-diagonal of the matrix $U_{n,m}$ which are of the order of $E_q$; the
elements in the second off-diagonal will be of the order of $(E_q)^2$
and so on.
Let us estimate the width $\Delta n_i$ of the maxima of the
separatrix eigenfunctions, where the index $i$ labels the
separatrix number. To this end we approximate the matrix element
$V_{n,n+1}$ near the separatrices by the linear function (see upper
part of Fig. 1) $V_{n,n+1}=V_{n_0,n_0+1}+\alpha(n-n_0)$. Here $n_0$
is the Landau state number where the value of the matrix
element $V_{n,n+1}$ is minimum, and $\alpha = \partial
V_{n_0,n_0+1}/\partial n$. This approximation is valid when the
separatix region $\Delta n_i$ is smaller than the total number of the
Landau states in the i-th cell $n_i$, $\Delta n_i\ll n_i$. Under this
condition, Eq. (\ref{algebraic}) takes the form
\begin{equation}
\label{separatrix_width} {E_q\over
V_0}C_n^q=[V_{n_0,n_0+1}+\alpha(n-n_0)]C_{n+1}^q+
[V_{n_0,n_0+1}+\alpha(n-n_0-1)]C_{n-1}^q.
\end{equation}
 For the separatrix states, the ratio $E_q/V_0$ is of the order
of $10^{-3}$ and one
can omit the left-hand side of Eq. (\ref{separatrix_width}); the
minimal matrix element $V_{n_0,n_0+1}$ near the separatrix is also
small and may be neglected as well. Under these assumptions,
Eq.(\ref{separatrix_width}) gives the following relation between the
coefficients $C_{n}^q$:
\begin{equation} \label{reccurence}
C_{n_0+m}^q=\frac{m}{m+1}\frac{m-2}{m-1}\dots\frac12C_{n_0-1}^q,
\end{equation}
 Here $C_{n_0-1}^q$ is the magnitude of the QE eigenfunction in the
maximum and $m$ (here odd) is the distance from the maximum
(see insert in Fig. 1).
The reduced equation (\ref{reccurence}) is independent of the
parameters and cell number;
in consequence, the separatrix
width $\Delta n_i$ should be the same for all cells.
This was confirmed by our numerical calculations
discussed below.
In the quasiclassical limit, when $h\rightarrow 0$ and
$n_i\rightarrow\infty$, the relative width of the quantum separatrix
$\Delta n_i/n_i$ tends to zero, consistent with the classical
dynamics.

To conclude our discussion on the structure of the QE eigenstates in
the resonance approximation it is necessary to point out that the
numerical results obtained by using two different approaches ---
the Floquet Hamiltonian formalism \cite{Reichl_book} in Eq.
(\ref{algebraic}) and the Floquet formalism at $V_0\rightarrow 0$
described in Sec. II --- lead to the same results with very high
accuracy. The latter method is nonperturbative and allows us to obtain
the solutions at any (not necessarily small) amplitude $V_0$. In the
following discussion this approach  will be used for investigating
quantum chaos in our system.

In order to incorporate chaos one must
increase the perturbation amplitude. The structure of the QE
eigenfunctions in the presence of weak chaos is shown
in Figs. 2 (b), and 2 (c). Note the qualitative difference
from the results obtained for
the resonance approximation (Fig. 2 (a)).
Increasing the perturbation amplitude to the value $V_0=6$ (Fig. 2
(b)) gives rise to a change in the separatrix eigenstates: their
dispersions $\sigma_q$ decrease on the average. In contrast,
localized eigenfunctions (arranged in rows) are slightly
affected by the perturbation, resulting in a small splitting
of the rows. This is in close agreement with the
classical behavior; namely, the region around the classical
separatrix is first and most affected by an increase of $V_0$.

Further increase of $V_0$
to the value $V_0=13$ (Fig. 2 (c)) leads to an increase in the
number of delocalized states, and on the average, the localization
length grows again. The most
drastic effect of chaos on the quasienergy eigenfunctions was
observed in the region of Hilbert space corresponding to completely
chaotic motion in classical phase space (the first two cells in
the right part of Fig. 2 (c)). The chaoticity of the corresponding QE
eigenfunctions is manifested in the apparent random character of the
dependence
$\bar n_q(\sigma_q)$; each QE eigenfunction spans both cells  so that
they cannot be assigned to any particular one. Furthermore,
our numerical experiments show that the dependence of $C_n^q$ on the Landau
number $n$ in the chaotic area is also very irregular. Note that,
as in  the classical case, perturbation affects the various cells differently: cells
with small values of Landau numbers $n$ appear to be more affected
than those with larger $n$.

In Figs. 2 (b), 2 (c) there are no points corresponding to two eigenfunctions
as there are in Fig. 2 (a) because of the substantial influence of the
nonresonant terms on all the QE eigenstates. Recall that in the resonance
approximation the system has the symmetry defined by Eq. (13). Weak chaos
lifts this symmetry and splits the rows on the left-hand side of
Figs. 2 (b) and 2 (c). The 3rd and 5th rows correspond to
mixed phase space
dynamics, shown on the right hand-side of Figs. 2 (b) and 2 (c).

\section{Quantum diffusion via the separatrices}
The quantum dynamical manifestations
of weak chaos are  studied in this section
by means of the nonperturbative
technique based on the Floquet formalism discussed in Sec. II.
In the regime of weak chaos
only a small portion of phase space is chaotic.
Correspondingly, in the quantum model, only a small number
of the eigenstates are affected (see Sec. III).
However, we expect to detect the influence
of weak chaos on the dynamics if we consider the diffusion along
the separatices. This intuition is based on the fact that
the diffusion via the separatrices is governed by the separatrix
QE eigenstates, which are the ones mainly affected by the perturbation.
In order to check this, let us consider the
evolution of an initial state $C_n^{n_0}(0)=\delta_{n,n_0}$, which is
described by (see Eqs. (\ref{Green_f}) and (\ref{Green_evolution}))
\begin{equation} \label{delta_dynamics}
C_n^{n_0}(M)=\sum_qC_n^qC_{n_0}^{q*}\exp(-iE_qM), \end{equation} where
time is measured in the number of external field periods $M$. The
value of the $q$-th oscillating term is determined by the $n$-th and
$n_0$-th amplitudes of the $q$-th QE eigenfunction $C_n^q$. A transition
from $n_0$-th to $n$-th Landau level will occur provided both
coefficients $C_n^q$ and $C_{n_0}^{q*}$ in Eq. (16)
are large. If we consider the transitions between the quantum
separatrices, then the main contribution to the evolution comes from
the separatrix QE eigenfunctions as they have their maxima at the
various separatrix regions, and the diffusion over the quantum
separatrices is due to their delocalization. The number of separatrix
QE eigenfunctions is very small but their effect on the diffusion via
the separatrices is crucial because localized QE eigenfunctions have 
minima in the vicinity of the separatrices and do not contribute to 
this process.

First, let us look at the diffusion over the quantum separatrices in
the resonance approximation where the system possesses no
chaos. To this end, we place an initial state $C_n(0)=\delta_{n,n_0}$
in the separatrix region and follow the evolution of this state for a
sufficiently long time in order to determine the position in Hilbert
space of the
quantum particle at time $t\rightarrow\infty$. The largest
characteristic time in the system is $t_{max}=2\pi/\omega_{min}$,
where $\omega_{min}$ is the minimal distance between the effective
QE eigenvalues, i.e., the eigenvalues corresponding to the eigenstates
which constitute the initial state and determine the dynamics. As shown
above, in our case the effective eigenstates are the
separatrix ones. After a time $t_{max}$ the dynamics enters into a
quasistationary regime and it is convenient to time-average the
probability in the region $t\gg t_{max}$ in order to eliminate the
influence of fluctuations and exclude  transient effects.

The result of this procedure is plotted in Figs. 4 (a) and (b)
for the same $h$ as in Fig. 2 and for two different values of
$V_0$; the initial state, marked with a large arrow, was
situated at the Landau level $n_0=20$ in the second cell near the
boundary of the first one. From Fig. 4 (a) one may note that the 
time-averaged
probability  distribution $\left < P_n \right >$ in the classically 
inaccessible cells (even
in the resonance approximation, where $V_0\rightarrow 0$) is
comparable with that of the initial (second) cell. It was numerically
confirmed that if an initial state $C_{n_0}(0)$ is situated anywhere
in the central region of a resonance cell, then only an exponentially
small part of a wave packet tunnels to the neighboring cells and a
logarithmic scale (in the probability distribution)
 is necessary in order to recognize the
tunneling phenomenon. \cite{2} The second principal point is the
observation that the probability distribution in Fig. 4(a) is highest
around the boundaries of the quantum resonance cells, being always
relatively small in the central regions. Thus, anomalously intensive
tunneling takes place only between the quantum separatrices, and we may
refer to this process as "diffusion via the quantum separatrices." This
effect is reminiscent of the diffusion of a classical particle within
the stochastic web. That idea is supported by the Husimi function of
the initially $\delta$-like wave packet after evolution during a
rescaled time $\tau=3000$ (see Fig. 5 (a)). This figure reminds us
of the web structure in its corresponding classical phase space.
However, this effect is of a pure quantum nature as the classical
particle in the resonance approximation has no possibility of
penetrating from one cell to another.

The effect of weak chaos on the diffusion over the separatrices
is illustrated in Fig. 4 (b).
Inspection of Figs. 4 (a)  and 4 (b) shows that an
increase of the amplitude leads to a decrease of
the diffusion rate. This effect of chaos on quantum
diffusion is rather unexpected: the increase in the perturbation
inhibits tunneling instead of intensifying it.
This is a direct consequence of the partial localization of the
separatrix QE eigenfunctions of Fig. 2 (b) caused by the
presence of weak chaos in the vicinity of the separatrix. This
quantum weak chaos effect is manifested in the plot of the Husimi
function of the wave packet $C_n(0)=\delta_{n,n_0}$ after evolution
during the time $\tau=3000$ ( see Fig. 5 (b)). Hereafter the time $t$
is measured in units of $T$. The third separatrix is
not as clearly defined as in Fig. 5 (a) or in Fig. 3, being partially
destroyed by chaos. The Husimi function looks like  the classical density
distribution within the stochastic web in phase space (see Fig.
5.7 in Ref. \cite{Zaslavsky}). The 4th and 5th separatrices are
absent and the Husimi function appears to be more localized than in
the resonance approximation. The structure of the Husimi function
within the first two cells is irregular, consistent with the more
developed (not weak) chaos in this region.

The phenomenon of localization becomes more evident in Fig. 6
where we show the evolution of the squared dispersion $\sigma^2$
as a function of the rescaled time
$\tau=V_0t$.
 The two chosen values of the perturbation
amplitude $V_0$ correspond, respectively,  to the resonance 
approximation and the
regime of weak chaos (the values of $h$ and initial conditions in
Fig. 6 are the same as in Figs. 4 (a), (b)). We see from Fig. 6  that
the diffusion over the separatrices at $V_0=6$ is suppressed in
comparison with the resonance case, consistent with the dynamical
picture of Fig. 4. The data shown in Fig. 6 allow us to estimate the
effective time of saturation of the probability distribution
$\tau_{max}=V_0t_{max}$. As the wave packet in Fig. 6 for $V_0=0.002$
spreads  over all the separatrices, the effective time in this case
is defined by the minimal distance between the separatrix QE
eigenvalues; in the presence of weak chaos ($V_0=6$) $\tau_{max}$
is mainly determined by the minimal distance between the QE
eigenvalues within the initial cell, because the probability
distribution does not evolve to the other cells. The separatrix QE
eigenvalues are situated near the center of the spectrum where the
spectrum is dense and the distances between these levels are smallest
\cite{1}. Thus, a large difference in the values of $\tau_{max}$ in
the two cases of Fig. 6 arises from the different types of
effective eigenstates which determine the dynamics, namely,
$\tau_{max}(V_0=0.002)\gg\tau_{max}(V_0=6)$.

 We will now show that the diffusion rate can be characterized by two 
parameters of the probability distribution $P_n$ in the neighborhood 
of each separatrix $i$. These are the maxima of $P_n$, denoted by $P_i$, 
and its width $\Delta n_i$.
Plots like Fig. 4(a) suggest that the distribution $P_n$ in the
neighborhood of the separatrix $i$, averaged over time, may be
approximated by a gaussian
curve. Then the maxima $P_i$ ( width $\Delta n_i$) is given by the 
height ( width) of the corresponding gaussian.  Plots of 
$P_i$ as a function of the perturbation amplitude $V_0$ give a 
measure of the diffusion rate. Inspection of Fig. 7 reveals an 
exponential decrease of the diffusion rate
up to the value $V_0'\simeq 6$ ($V_0'$ depends on $h$). This provides
further evidence of the suppression of quantum diffusion over the
separatrices due to weak chaos. Further increase of $V_0$ results in a
growth of the diffusion rate which can be explained by the average
increase in the localization length discussed in Sec.III. Such
behavior as a function of amplitude corresponds to the classical
situation. The minimum in the curve $V_0'$ is the crossover
from the quantum to the classical diffusion.  The further the
boundary of the resonant cell is from the boundary where the initial
state was situated, the smaller the minimum is in the corresponding curve
because chaos destroys a larger number of the delocalized
separatrix eigenfunctions which contribute to the diffusion. We
speculate that the oscillations of the curves in Fig. 7 arise from the
fact that the diffusion over the quantum separatrices is determined by
a small number of (separatrix) QE eigenstates, and turn out to be
very sensitive to changes in the structure of the QE eigenfunctions.

  Now consider the plots of the widths $\Delta n_i$ versus the wave
amplitude $V_0$  shown in Figs. 8 (a), (b), and (c) for the same
parameters, initial conditions, and separatrices as in Fig. 7. The
data, presented in Fig. 8, have the following interesting features: (i)
Consistent with the above considerations (see Eq.
(\ref{reccurence})), the widths of the quantum separatrices at
$V_0<V_0'$ are approximately the same for all the separatrices and for
all values of $h$ (the latter was confirmed by our calculations
with other values of $h$). (ii) On the average, the widths $\Delta
n_i$ do not change significantly with $V_0$ until the amplitude of the
perturbation reaches some value $V_0'$, after which the $\Delta n_i$ begin
to grow monotonically. Thus, in the region $V_0<V_0'$, classical
chaos does not affect the width of the quantum separatrices (but does
affect the diffusion rate in Fig.. 7). (iii) The threshold $V_0'$ in
the plot of $\Delta n_i(V_0)$ is the same as the threshold (position
of the minimum) in the plot of $P_i(V_0)$ in Fig. 7. At that point the
width of the quantum separatrix exceeds the width of the classical
stochastic web $\Delta H_i$ which can be approximated by
\cite{{Zaslavsky}} 
\begin{equation}
\label{classic_separatrix_thickness} {\Delta H_i\over \hbar\omega_c}=
2^{1/2}\pi^{7/2}{(2hn_i)^{1/2}\over V_0h^2}\exp\left[ (-\frac\pi
2)^{5/2}\frac{(2hn_i)^{1/2}}{V_0h}\right], 
\end{equation} 
where $n_i$
is the center of the $i$-th cell and the quantity $hn$ in the
quasiclassical limit becomes the action $I$. The results of our
calculations using Eq. (\ref{classic_separatrix_thickness}) are
presented in Figs. 8 (a), (b), and (c) with  dashed lines.
The discrepancy
between the quantum and classical curves may presumably be
attributed to the approximate character of Eq.
(\ref{classic_separatrix_thickness}) which is valid only in the case
of an  exponentially thin separatrix. Nonetheless, the trend of the quantum
curves is qualitatively the same as the classical behavior.

\section{Conclusion}
The numerical data and qualitative analysis presented in this paper
allow us to make the following conclusions about the nature of quantum
weak chaos. In the quantum resonance approximation we investigated
the resonance structure of Hilbert space and also the new phenomenom
of quantum diffusion via the separatrix. We remark that this quantum
diffusion has no classical analog because classically, the orbits
are confined to resonance cells. In quantum mechanics,
this diffusion results
from tunneling across the quantum separatrices.
Accidental intersection of levels of different cells
(associated with the overlap of QE functions)
with  eigenvalues $E_q\ll V_0$ leads to the formation of  delocalized QE
eigenfunctions. In other words, the cell structure of the evolution
operator matrix (and the Floquet Hamiltonian matrix) gives rise to a long
range coupling between states of different cells. The dynamical
manifestation of this effect is an anomalously large diffusion rate
between the cells via the quantum separatrices. 
For sufficiently large values of $V_0$ the nonresonant
terms may be considered as an effective random perturbation
which inhibits the long range
interaction, thereby localizing the QE eigenstates.

When the perturbation becomes strong enough diffusion is recovered,
but now it is of a completely different nature.
Namely, for large
values of the perturbation we observe an analog of the classical
diffusion within the stochastic web. This  is demonstrated by comparing
the widths of the
classical and quantum separatrices. The structure of the quasienergy
eigenstates which explains the  diffusion is also different: in the
resonance case they have the regular form, which is maintained
over several cells of Hilbert space, while at
large $V_0$ such states
are destroyed and diffusion takes place because the localization
lengths of many QE eigenstates increase on the average.

In this paper the evolution operator propagating the system toward an
arbitrary number of periods of the external field is built in the
Hamiltonian $\hat H_0$ basis. Its eigenstates (QE eigenstates) are
explored under the condition of resonance in the regime of weak
chaos. In
the new phenomenon of  diffusion over the quantum separatrices, it was
found that  a small number of delocalized separatrix QE eigenstates
play the dominant role. It was shown that weak quantum chaos leads
to the localization of the separatrix eigenstates and, hence, to
suppression of  quantum diffusion via the separatrices. At large values of
perturbation, $V_0>V_0'$, we have observed a recovery of the diffusion
which was associated in the quasiclassical limit with the growth of
the classical stochastic web in phase space.

It is necessary to point out that the parameters chosen in our
numerical experiments correspond to actual experimental situations. Acoustic 
cyclotron resonance can be observed
in a 2D electron gas in semiconductor
heterostructures subject to a transverse magnetic field, and in the
field of a longitudinal sound wave. In order to observe this phenomenon 
the electron
relaxation time $\tau_p$ must be large enough to satisfy the inequality
$\omega_c\tau_p\gg 1$.
Under this condition, and under the condition of cyclotron resonance,
one can choose parameters which allow the Fermi level $n_F$ to
be placed at the boundary (quantum separatix) between the first and 
second cells. That is, the argument of the Bessel function 
$J_1(ka\sqrt{2n_F})$
must coincide with the first zero
of the Bessel function $J_1$.
In order to create this situation in an experiment one can choose the
following experimental parameters: the sound wave frequency should be of the
order of 10 GHz, the magnetic field
$H=2*10^{3}$ Oe, the effective electron mass  $m^*=0.7m_e$,
and the electron concentration $\aleph=10^{11}$ cm$^{-2}$.
These parameters give:
$h=0.4$, $n_0=n_F=20$.
The value of the parameter $V_0$ is determined by the deformation
div({\bf u})$\sim ku_0$, where $u_0$ 
is the acoustic wave amplitude. Thus,
the value $V_0=10$ corresponds to the wave deformation $ku_0\sim 10^{-4}\div 10^{-5}$.
When the proposed parameters are
realized in an experiment we predict that quantum chaos will be manifested
as, for example, an attenuation of the sound wave.

\section{Acknowlegements}
We thank Felix Izrailev for illuminating discussions.
This work was performed with the support of INCAS
(Grant No. 97-2-15) and the Russian Foundation for Basic
Research (Grants No. 98-02-16412 and 98-02-16237). 
G.A.L-A acknowledges partial support
from CONACYT(Mexico) grant, No. 26163-E.

\section{Figure captions}
\begin{figure}
\caption{The matrix elements $V_{n,n+1}$ in dimensionless units versus
the Landau number $n$ (upper part); the most delocalized QE
eigenfunction in the resonance approximation (lower part). The
insert amplifies a small portion of the QE eigenfunction, marked with
brackets. Here $h=0.37$, $V_0=0.002$} \label{Fig. 1} \end{figure}

\begin{figure}
\caption{Plot of the dispersions $\sigma_q$ versus $\bar n_q$ for the
QE eigenfunctions with $q=1,\,2,\,\dots,\,N$  for $h=0.37$, $N=381$
and different values of $V_0$. (a) $V_0=0.002$, (b) $V_0=6$, (c)
$V_0=13$ (left hand-side); classical phase space for the same
parameters (right hand-side).}
\label{Fig. 2}
\end{figure}

\begin{figure}
\caption{ The Husimi function corresponding to the QE eigenstate
plotted in Fig. 1.} \label{Fig.3}
\end{figure}

\begin{figure}
\caption{The time-averaged probability distribution $\left < P_n \right >$
versus $n$, $h=0.37$, $N=381$, (a) $V_0=0.002$, (b) $V_0=6$.
Averaging was performed over $300$ times in the region
$3\,000<\tau<6\,000$ ($\tau=V_0t$; t is measured in units of
the external field period $T=2\pi /\omega$).
The separatrix positions are marked with arrows.}
\label{Fig. 4}
\end{figure}

\begin{figure}
\caption{The Husimi functions of the states for
the same parameters and initial conditions as in Fig. 4
at (dimensionless) time $\tau=3000$, $C_n(0)=\delta_{n,n_0}$,
$n_0=20$, (a) $V_0=0.002$, (b) $V_0=6$.} 
\label{Fig. 5} 
\end{figure}

\begin{figure}
\caption{The dynamics of the squared dispersion $\sigma^2(\tau)$  for
the same parameters and initial conditions as in Figs. 4(a) and 4 (b)
in units of dimensionless time.} 
\label{Fig. 6} 
\end{figure}

\begin{figure}
\caption{The diffusion rate $P_i$ of an initially $\delta$-like
wave packet, placed at $n_0=20$, from the initial separatrix (first)
to the i-th one for $i=3,\,4\,,5$ as a function of the perturbation
amplitude $V_0$. $h=0.37,\, N=401$.}
\label{Fig. 7} 
\end{figure}

\begin{figure}
\caption{The widths of the quantum separatrices $\Delta n_i$
versus the perturbation amplitude $V_0$ for the same parameters, 
initial conditions and separatrices as in Fig. 7.} 
\label{Fig. 8}
\end{figure}

\end{document}